\documentstyle[aas2pp4,psfig]{article}

\newcommand{\be}{\begin{equation}}
\newcommand{\ee}{\end{equation}}
\newcommand{\nn}{\mbox{} \nonumber \\ \mbox{} }
\newcommand{\ba}{\begin{eqnarray}}
\newcommand{\ea}{\end{eqnarray}}
\newcommand{\om}{\omega}

\begin{document}
\title{GRBs from unstable Poynting dominated outflows}
\author{Maxim  Lyutikov}
\affil{Canadian Institute for Theoretical Astrophysics,
 St George Street,
Toronto, Ontario, M5S 3H8, Canada}
\author{Eric G. Blackman}
\affil{Dept. of Physics \& Astronomy,  and Lab. for Laser 
Energetics, University. of Rochester, Rochester NY, 14627, USA}

{\bf Key words:} radiation mechanisms: non-thermal: general - gamma-rays: bursts.

\begin{abstract}
Poynting flux driven outflows from magnetized rotators 
are a plausible explanation for  
gamma-ray burst  engines.  We suggest a new possibility for
how such outflows might transfer energy into radiating particles. 
We argue that the Poynting flux drives non-linearly  unstable
large amplitude electromagnetic waves (LAEMW) which 
``break''  at radii  $r_t \sim 10^{14} $ cm where the MHD
approximation becomes inapplicable. 
In the ``foaming'' (relativisticly reconnecting)
 regions formed during the wave breaks the random electric fields
stochastically  accelerate particles to ultrarelativistic energies
which then radiate in  turbulent  electromagnetic fields.
The typical energy of the
emitted photons is a fraction of the 
fundamental Compton energy $ \epsilon \sim f  \hbar c/r_e $ with 
$f \sim 10^{-3}$ plus additional boosting due to the bulk motion
of the medium.
The emission properties are similar to synchrotron radiation, 
with a typical  cooling time $\sim 10^{-4}$ sec.
During the wave break, the plasma is also bulk accelerated in 
the outward radial direction and at larger radii can produce 
afterglows due to the interactions with external medium.  
The near equipartition
fields required by afterglow models maybe due to 
magnetic field regeneration in the outflowing plasma (similarly to the  
field generation by LAEMW
 of laser-plasma interactions)  and mixing with the upstream plasma.
\end{abstract}

\section{Introduction}

A Poynting flux driven outflow from a magnetized rotator
is a promising paradigm for gamma-ray burst (GRB) engines and 
there have been various implementations of this concept (c.f. Usov 1992;
Thompson 1994; Blackman et al. 1996; M\'esz\'aros \& Rees 1997;
Kluzniak \& Ruderman 1998).
Such models are appealing in light of the fact that 
neutrino driven GRB emission from compact object mergers
(Ruffert \& Janka 1998; Janka et al. 1999)
might fall short in radiative luminosity. Also, Poynting flux
models may provide a source of magnetic fields
which could help alleviate 
the  field amplification problem in GRBs themselves and in afterglows which requires
fields  of the order of equipartition
(\cite{WG98,FWK99,GPS99}).

Scenarios that could produce  Poynting flux
 dominated outflows (PFDOs) require a source
of magnetic fields  $\ga 10^{15}$ Gauss,
and rotation speeds of order $\Omega \sim 10^4$/sec.
The total available energy from a  compact ($R_0 \sim 10^6$ cm, 
$M \ge M_{\odot}$) object is then
$ M \Omega^2 R_0^2/2 \ge 10^{53}$ ergs and a dipole type luminosity
$ 2 B_0^2 \, R_0 ^6 \,\Omega^4/ 3 c^3 \sim 2 \times 10^{50}$ ergs/sec.
Such combinations could be generated in a number of plausible cases: 
an accretion torus surrounding a black hole that 
formed from a neutron star merger 
(Rees \&  M\'esz\'aros 1997), strongly magnetized neutron stars (``magnetars''
Duncan \& Thompson 1992) possibly formed from accretion induced collapse of a
white dwarf with possible dynamo amplified fields in the hot young 
neutron star (Usov 1992; Duncan \& Thompson 1992,  
Blackman et al. 1996), failed supernova Ib  
(Woosley 1993),  rapidly rotating magnetized 
black holes undergoing 
a Blandford-Znajek type energy extraction from supernovae
(Lee et al. 1999,Brown et al., 2000)
or hyper-accreting accretion disks (Popham et al. 1999),
with an MHD  dynamo (Araya-Gochez 2000).


Without further specifying the nature of the
central engine we suppose 
that GRB are due to such 
stellar mass objects, so that the  conditions  required for the
generation of PFDOs 
are satisfied.
Here we are interested in how Poynting flux models accelerate particles
and produce gamma-ray emission.

The suggestion that GRBs are PFDOs 
whose energy is converted to particles and escaping radiation only at large 
distances from the central rotator
potentially accounts for some important characteristics of GRBs (c.f.
Blackman et al. 1996, Rees \& 	 M\'esz\'aros 1997) 
(i) mass loading -  PFDO can  be launched without much matter,
and particularly without much baryon contamination 
(as in pulsars, the wind may carry no baryons at all);
(ii) compactness problem -
the PFDO converts its energy effectively into particles
and gamma-rays at distances many orders of magnitude from the 
central engine, 
(iii) collimation -   PFDO have a preferred direction along the rotation
axis of the central engine;
(iv) jets  observed in pulsars and (probably) AGNs provide examples of operation of PFDO;
(v) 
the energy from Poynting flux can be easily converted into high frequency
electromagnetic radiation.  The presence of strong magnetic 
fields from the rotator provides  electromagnetic
energy which might be tapped or converted into the fields 
which accelerate the radiating particles.

PFDOs may differ from the conventional blast wave models in several 
important respects.
The lepto-baryonic content of the outflow in the inner region
is expected to be small by analogy to pulsar
winds, and the energy carried by matter in the central region 
will be less than the energy carried by 
EM fields. As in the relativistic pulsar wind, the only other well established
PFDO,  the baryons may be absent altogether while energy flux in the
electron-positron pairs may be as small as one millionth of the  
Poynting flux. 

The conventional estimates of the relativistic expansion Lorentz factors 
and density of the outflowing plasma used in the framework of blast wave 
models may not be applicable to the Poynting dominated outflows
because of the strong magnetic fields (Paczy\'{n}ski 1990, Usov 1999). Strong fields 
dramatically reduce the importance of pair
production from photon-photon interactions 
relative to the photon-magnetic field  interactions.  
Other exotic QED processes, like photon splitting 
(e.g., Baring 1991, Adler 1971),
will also become important for magnetic fields larger than the critical value 
of $4.4 \time 10^{13}$ Gauss. 
At this moment, we are unaware of the calculations of the Compton scattering 
cross section in supercritical magnetic fields, but we expect that it will 
also be strongly suppressed.

The above suggests that  in the central regions of  PFDOs, 
the magnetic pair production and
photon splitting  may be more
important than pair production in photon collisions. 
Note that this situation 
is realized in pulsars. There acceleration of particles to energies 
beyond $10^6$ MeV proceeds relatively quietly, 
without producing a large number of observable photons.
Very few rotationally powered  pulsars are observed at 
high energies. Thus conventional estimates of the Lorentz factors and 
densities  in GRBs based on two  photon pair production  may be irrelevant.
 
To estimate a ``revised'' density of the generated pairs in the strong
magnetic rotator case applicable to GRB, we use 
the analogy with pulsars where the number of pairs produced is related to the
local Goldreich-Julian density $n_{GJ} = \Omega B /( 2 \pi  e c)$ 
(Goldreich \& Julian (1969)).
The pair number density 
 is usually $\lambda \sim 10^3 -10^5$ times larger than $n_{GJ}$ in
the inner regions. The factor $\lambda$
depends sensitively on the magnetic field, 
its curvature and accelerating potential. A
highly curved magnetic field produces denser and less relativistic
outflows.

In this  paper we  address
the fundamental but  unresolved issue of how 
Poynting flux is converted into gamma-rays and the subsequent
emission characteristics.
In section 2 we give the basic picture of the magnetized rotator
magnetosphere and the propagation of large amplitude electromagnetic
waves. In section 3 we discuss the breaking of these waves, and 
the predicted emission from random electromagnetic fields is discussed
in section 4. There we address the spectrum and time dependence of the emission. The afterglow possibilities from our picture are addressed in section 5. 
We conclude in section 6.

\section{Basic Picture}

Here we discuss the basic proposed scenario for the evolution of PFDOs.
If there is a strong overall magnetic dipole component, then
 near the central engine the  PFDO will resemble a pulsar. Along the 
magnetic poles it will produce jet-like flow
with  helical magnetic field, oscillating on  scale length of
 $ (c/\Omega )\sim 10^6$ cm. 
 Initially Poynting  energy flux dominates over the particle flux by a ratio
$\sim 10^6$ - typical for pulsars.

Inside the "light cylinder" $r_{LC}= (c/\Omega ) = 3 \times 10^6$  and the
magnetic field falls off as $r^{-3}$, while 
beyond the light cylinder
the magnetic field is  dominated by the toroidal component 
which falls of as $r^{-1}$. Thus at large distances $r \gg r_{LC}$ we have
\be
B_{\phi} = B_0 {R_0 \over r} \left( {\Omega R_0 \over c} \right)^2 
\ee
where $R_0 =10^6$ cm and $ B_0 =10^{15}$ G are the size of the central engine and
the initial magnetic field.

Following the analogy to pulsar winds (e.g. Usov 1992; 1994), 
at small distances the magnetic field of the wind is frozen in the plasma.
As the plasma flows out, the plasma density
decreases in proportion to $r^{-2}$ reaching a radius $r_t$, where
it becomes less than the  critical charge density $n_t$ required for 
applicability of MHD (Goldreich \& Julian 1969;
Michel 1969, 1994;  Coroniti 1990; Melatos \& Melrose 1996).
Locally  the critical charge density $n_t$ is equal to the  Goldreich-Julian 
density $n_{GJ}$. The latter decreases only as $r^{-1}$. 

In what follows, we normalize the real density of the 
PFDO $n$ to the  Goldreich-Julian 
density  at the light cylinder $ n_{GJ} (r_{LC})$, that is 
\be
n  = \lambda \, n_{GJ}
=\lambda { \Omega B_0 \over 2 \pi e c} \left( { R_0  \Omega \over c} \right)^3 =
4 \times 10^{15} {\rm cm ^{-3}}
\ee
 Since the plasma density satisfies 
$ n(r) = n(r_{LC}) (r_{LC}/r)^2$, while 
$n_{GJ}(r)=n_{GJ} (r_{LC}) (r_{LC}/r)$,
the plasma density equals Goldreich-Julian density at 
\be 
r_t = \lambda r_{LC} = 10^{14} {\rm cm}, 
\ee
where we have chosen $\lambda= 3 \times 10^7$ (see also Usov 1994; Blackman
\& Yi 1998).  The density and magnetic field at the transition radius
are 
\be 
n_t = n_{GJ} (r_t) = 10^8 {\rm cm^{-3}},
\hskip .2 truein 
B= 10^6  {\rm  G}.
\ee
The physical parameters near the transition region are
listed in Table \ref{Table}.

\begin{table}
\caption{Physical parameters near the transition region}
\begin{tabular}{cc} 
$ \phantom{ {{a\over b} \over {a\over b}}}  B, Gauss $ & $ 10^6 $  \\    
$\phantom{ {{a\over b} \over {a\over b}}}
 \om_B$, rad/sec & $ 2 \times 10^{13} $ \\ 
$ \phantom{ {{a\over b} \over {a\over b}}}  \nu_0 =
 {\om_B \over \Omega} $ & $ 2 \times 10^9 $ \\   
$\phantom{ {{a\over b} \over {a\over b}}}  
r_L = {c\over \om_B} $, cm & $ 1.5 \times 10^{-3} $ \\  
$\phantom{ {{a\over b} \over {a\over b}}} 
\om_p = \sqrt{2 \Omega \om_B}  $, rad/sec & $  6   \times 10^8   $ \\  
$\phantom{ {{a\over b} \over {a\over b}}} 
r_s = {c \over \om_p} $, cm & $  50  $ \\   
$\phantom{ {{a\over b} \over {a\over b}}} 
{r_L \over r_s} = {\om_p \over \om_B} =  
\sqrt{ { 2 \over \nu_0 }} $ & $ 3 \times 10^{-5}  $ \\  
$\phantom{ {{a\over b} \over {a\over b}}}  
{r_L \over r_e} $ & $ 5 \times 10^9 $ \\ 
\end{tabular}
\label{Table}
\end{table}

At $r>r_t$ the the wind field is transformed into a 
Large Amplitude Electromagnetic Wave (LAEMW). 
\footnote{Though formally the plasma at transition point is overdense, 
$\om_p > \Omega$ a presence of a small "external" magnetic field with strength
larger than $10^{-3} $ Gauss will allow wave propagation.}
The dimensionless 
strength parameter of the LAEMW, $\nu_0 = { e E /( m c \Omega)}$,
near the transition point is
 \be 
\nu_0 =\frac{\Omega \,{{R_{0}}}^3\,{B_{0}}}{c^2\,r_t} \sim 10^{9}.
\ee
This large value implies that the 
electromagnetic waves can drive particles to superrelativistic energies 
in one period of oscillation.  We suggest that gamma-rays  will be emitted 
at the point where MHD breaks down due to the 
overturn instability of LAEMWs.

The interaction of LAEMW with the plasma bears 
similarity to laser-plasma interactions. 
It is well known from laboratory experiments that 
interaction of  strong electromagnetic
radiation with plasma leads to a number of violent instabilities 
(e.g. Kruer 1988).  The most important instabilities include resonant 
excitation of plasma waves when the frequency of the  laser light matches the plasma frequency, and nonresonant ponderomotive instabilities.
The fact that GRB outflows may consist of purely electron-positron 
plasma might seem to play a stabilizing role, 
however pair plasmas are  also susceptible 
to parametric and modulational instabilities (Lyutikov \& Green 2000).
Here we point out  that in addition to  parametric instabilities 
 of EM waves, there is also an overturn instability 
associated with the 
non-linear evolution of the wave packet as it propagates through the plasma.   
We describe this  instability in the next section.

Once the EM  
wave overturns, it  creates a broad spectrum of random EM  fields. It is an 
electromagnetic analog of the foam of the deep water surface waves. 
Overturn of the initial coherent wave will create random EM fields and
currents which will be strongly dissipated.  In general,
 there will be a nonzero component of electric field along magnetic field. 
 We will call this state a relativisticly strong  electro-magnetic turbulence. 
Since the end result is the dissipation of the magnetic energy, this regions may also be called a
relativisticly strong reconnecting cites.
Particles will be
 accelerated in this EM turbulence to relativistic energies
(in the rest frame of the foam). 
As the  accelerated relativistic particles 
 move   through the EM  foam they generate the
observed GRB emission. The emission is due to the acceleration in the
random E-B fields.  
Emission would emanate from multiple 
breaking regions.  The separate spikes in the GRB profiles  would be 
due to separate "foaming regions" - separate regions of the wave 
break.   

In many respects the  "foaming regions"  resemble optical shocks 
observed in the laboratory when an optical pulse steepens as it  
propagates in an optically active medium (Agrawal 1989).  
A wave overturn creates an electromagnetic shock.
Interaction of such EM shocks will  unavoidably lead to a creation of internal
 plasma discontinuities with  strong dissipative effects. 
The "foaming regions"  play the role of 
``internal shocks'' (c.f. Piran 1999 for a review) of GRB theory.   
This term is usually used to mean ``internal processes
in the blastwave which generate bursts'' not necessarily a specific
mechanism, which we are providing here.
Alternatively, the "foaming regions" maybe thought of as relativisticly 
reconnecting regions, where the energy of magnetic field is converted into 
particles in a relativisticly turbulent manner.

It is often assumed that some kind of Fermi scattering
process operates at the GRB relativistic shocks.
In the shock Fermi process, particle scattering occurs across the shock
between ``scattering centers'' of collections of e.g. Alfv\'en waves.
Technically speaking, shock Fermi acceleration is not the likely
mechanism for particle acceleration for generic relativistic shocks
because for upstream magnetic field angles greater than $1/\Gamma$ to the shock
normal, where $\Gamma$ is the bulk upstream Lorentz factor, the particles
would have to travel faster than $c$ to get back upstream along field lines.
Thus unless the shock is strictly parallel (unlikely situation) then
shock Fermi acceleration won't work.

Hoshino et al. (1992) studied 1-D models of 
positron acceleration in relativistic perpendicular shocks (where
the field is perpendicular to the shock normal) in ion-electron-positron
plasmas. Their mechanism involves the absorption, by positrons, of gyro-phase
bunched magnetosonic waves emitted by protons reflected at the shock front.
For $\sim 10$\% ion fraction, non-thermal tails are seen in positrons.
They discuss limitations of their simulations 
with respect to their choice of 
electron-position number ratio, the fact that the
simulations are 1-dimensional, and that they do not consider 
oblique shocks (shocks for which the fields are not exactly perpendicular to the shock normal).  They point out that electric fields may play 
more of a role in particle acceleration in the latter case. In general, 
they agree that more work is needed.   

We entertain the possibility that electric fields may be important
for acceleration, but not necessarily only across a thin shock.
Rather, we suggest that as an LAEMW overturns, 
the acceleration, even in pair plasmas, might be 
done directly by turbulent electric fields (see also Lembege \& Dawson 1987
for overturn discussion).  The thickness of the turbulent region
may be much larger than the thickness of a typical shock.
The overturn may force the region to be highly strongly turbulent over
many gyro-radii. 

The picture of the "foaming regions"  bears some similarity to the 
"plasma turbulent reactor"  (Tsytovich \& Chikhalev , 
Norman \& ter Haar 1975; 
Rees  1967). In both cases there is a strong  interaction 
of plasma kinetic turbulence with
electromagnetic waves which may lead  to a formation  of stationary 
distribution
of particles.  But the "plasma turbulent reactor"  has been studied
only in the limit of weak turbulence theory, 
when the energy density of the turbulent  plasma fluctuations
is much smaller than the total energy density of plasma and when 
plasma is optically thick to transverse EM waves. 
 The assumption of weak turbulence allowed for a closed  
self-consistent solution  
for the plasma turbulent spectrum and particle distribution. 
On the other hand, 
the turbulence in the foaming regions of GRBs should be very strong, 
with an energy density much larger than that of the plasma.
Detailed properties of such superstrong  turbulence such as the stationary 
spectra, are not known. Under certain
conditions we will show that the emission from such  turbulence does not depend
in on its detailed properties (see section 4).
We will not restrict ourselves to a particular model of the
turbulence, but 
we do discuss Phillips-type and relativistic hydrodynamics-type spectra
in section 4.4 and 4.5.

Each overturning region produces a burst of  gamma-rays.
The temporal behavior of such a  burst might be most similar to a FRED 
(fast rise exponential decay) curve, since the fraction of 
LAEMW energy going into turbulence and particle acceleration likely occurs more
rapidly than the subsequent radiation drain. 
The overall duration of a microburst is of the order of micro seconds
(Eq. (\ref{tau_r})). To account for the energetics ($\epsilon_{mb} 
\approx 10^{48} $ergs per microburst with of order $10^4$ microbursts
as discussed later in section 4)
the overturning region should draw energy from a volume 
$V \sim 10^{36} {\rm cm}^3$, i.e. the typical size of the turbulent region 
should be $\sim 10^{12}$ cm  which is  much smaller
than the distance to the central region.  Two other arguments may influence our estimates
of the emitting region: (i) using the analogy with surface foam
we can expect the transverse dimensions of the foaming region
to be larger than the longitudinal dimension, (ii) 
during a break of a wave packet energy from many wave periods is
"piled up"
in a narrow breaking region, reaching the thickness of one wavelength $\sim 10^6$ cm. 
This value also turns out to be of the order of 
$c$ times the radiative decay length of  a typical
particle.

The total number of  overturning regions will depend on the structure of the
flow, which is in turn determined by the structure of the central engine. 
For example, 
if the PFDO has  a large dipole component, then the structure of the flow
will be more  regular and we expect only several regions of wave overturn. On the
other hand if the structure of the magnetic field near the central engine
is more multipolar,  we expect many "foaming regions".

At the transition radius, momentum conservation
dictates that there will remain some energy
in the LAEMW which can be used  for bulk plasma acceleration. 
The bulk motion of the outflowing plasma should drive
a relativistic shock wave into the ambient gas. 
Synchrotron emission of 
particles at this shock could produce the observed afterglows. 
This would coincide with the conventional external shock
models of afterglows (c.f. Piran 1999), while providing 
a possible improvement: 
as we discuss further in Section 5,  the  outflowing plasma beyond the
 transition radius  may still be strongly magnetized, thus 
supplying the elusive magnetic field
required in the synchrotron  models of afterglows 
\footnote{In the turbulent
region  magnetic fields with scales smaller than $\sim 10^6$ cm
 are annihilated, leaving (or regenerating)  larger
scale fields which obey frozen-in condition.}.
In this case the
 afterglows are something like the 
the analogs of the  wisps observed around the Crab pulsar
(the wisps are thought to be the location of the  reverse shock where
the pulsar winds pressure is balanced by the pressure of the supernova ejecta; 
Gallant \& Arons 1994).
Alternatively,  the afterglows may be generated
in a process similar to the one considered by Smolsky \& Usov (2000),
when the particles from the external 
medium are reflected from the strong magnetic field
of the ejecta.  In both cases however, we point out that
the outflow can supply the magnetic
field to the afterglow emitting region.

\section{Overturn of LAEMWs}

Here we consider the transition of the PFDO from MHD to 
the wave regime.  In the MHD approximation, the velocity of the wind equals  
the velocity of the field patterns. This wind velocity 
can be described as a wave phase velocity with trapped particles. 
In the laboratory  frame this velocity is determined by the electric 
drift in the crossed electric and magnetic fields, 
${\bf v}_d = { {\bf E} \times  {\bf B} \over  B^2 } = { E \over B} < 1$.
For the PFDOs, the leading  EM terms are 
$B_{\phi}$, $E_{\theta} \propto {1\over r}$ so that  the drift velocity
(the velocity of the outflow) 
 is directed along the radial direction ${\bf r}$.
When the MHD approximation breaks down, the phase velocity
(i.e. the velocity of the   field pattern) will grow to $ v_{ph} =
 { \Omega \over k} 
> 1 $. Thus, during the transition from MHD  approximation to
plasma wave solution the phase velocity of the field patterns will 
increase from subluminal to
superluminal. For plasma particles this process will
 look as if the EM wave is coming into plasma "from
minus infinity". 
As the EM wave penetrates into the 
plasma it exerts ponderomotive force along the
direction of its propagation which  provides an acceleration
for the plasma.
Below we show that this process of acceleration is unstable toward wave overturn.


The equations describing the evolution of the LAEMW waves in plasma are 
Maxwell's equations and
the mass and momentum continuity equations for electron and positron fluids.
To show the wave overturn simply, 
we consider a cold plasma. This is a reasonable approximation
when the particles acquire relativistic motion
under the influence of the wave.
 
Introducing the electromagnetic vector and scalar 
potentials ${\bf A}$ and 
$\phi$, in the Coulomb gauge ${\rm div } {\bf A} =0$ the equations become
(we use a system of units in which $c=m=1$)
\ba
&&
\left( \Delta -  \partial_t^2 \right)   {\bf A} = 4 \pi (n_p {\bf v}_{p} -
n_e  {\bf v}_{e} ) + \partial_t \nabla  \phi 
\nn
&&
 \Delta \phi = - 4 \pi e (n_p  - n_e )
\nn
&&
\partial_t {\bf p}_{e,p} + ( {\bf v}_{e,p} \cdot \nabla) {\bf p}_{e,p} =
\nn
&&
\hskip .2 truein \pm e \left(  \nabla \phi  + \partial_t  {\bf A} -
  {\bf v}_{e,p} \times \left(  \nabla  \times  {\bf A} \right) \right)
\nn
&&
\partial_t n_{e,p} + \nabla \left(   n_{e,p} {\bf v}_{e,p} \right)=0.
\ea 

The force equations  can be cast in the form
\ba
&&
\partial_t \left(  {\bf p}_{e,p} \mp e {\bf A} \right) -
{\bf v}_{e,p} \times \left( \nabla  \times   \left(    {\bf p}_{e,p} \mp e {\bf A} \right) \right) =
 \nn
&&
\hskip .2 truein
\left( \pm e \nabla \phi  -  \nabla \gamma \right)
\label{Vor}
\ea
with the immediate consequence that the generalized vorticity,
${\bf \Omega}=  \nabla \times \left( {\bf p}_{e,p} \mp   e {\bf A} \right) $
is conserved. If the plasma were initially 
quiescent without external field or vorticity, then
condition (\ref{Vor}) guarantees that the vorticity vanishes at all times.
Then  (\ref{Vor}) simplifies to 
\be
\partial_t \left(  {\bf p}_{e,p} \mp e {\bf A} \right) 
= \left( \pm e \nabla \phi  -  \nabla \gamma \right).
\ee

Consider the propagation 
of a strong plane electromagnetic wave  along the $z$ direction 
with a slowly varying amplitude ${\bf a} (t,z)$:
\be
 {\bf A}= {\bf a} (t,z)\, e^{-(\omega t - k_z z)}
\ee
We assume that variations of the amplitude occur on  the time scales much 
longer than the wave period and expand
the dynamic equations in small quantities 
${ \partial_t  a_{\perp} \over  \omega a_{\perp}} $ and
${\partial_z  a_{\perp} \over  k_z  a_{\perp}} $. In  nonlinear
optics this is known as a slowly varying
envelope approximation (e.g. Agrawal 1989).
To further simplify the treatment, we consider a 
circularly polarized EM wave for which
the energy of particles in a wave remains constant during oscillations   
 (polar outflows from magnetized rotator should produce 
circularly polarized EM waves). 
For linearly polarized waves, an additional averaging over one 
period of oscillation is needed. In the zeroth order, we find that 
\be
{\bf p_{\perp} }_{e,p} =  \pm  e \, {\bf a}_{\perp}. 
\hskip .2 truein p_z = {\rm const}
\ee
Thus the transverse velocity of the  plasma electrons and 
positrons are antiparallel and directed
 along the instantaneous magnetic field, but perpendicular to the electric 
field of the wave.  The longitudinal velocity is arbitrary.

To first order in the small parameters we find
\ba
&&
a_z = 0,  \hskip .2 truein \phi =0
\nn 
&&
\partial_t {p_z} +  \partial_z \gamma=0,  \hskip .2  truein 
\gamma = \sqrt{1+ \nu_0^2 + p_z^2}
\label{pz}
\\
&&
\partial_t n + \partial_z ( n v_z) =0
\label{n}
\ea
and a Schr\"{o}dinger type equation for the transverse vector potential
(e.g. Berezhiani et al. 1992).
Note that in the  unmagnetized   pair plasma, the 
variation of amplitude along the direction
of wave propagation does not create charge separation, so $n\equiv n_{e,p}$ and
$p_z \equiv {p_z} _{e,p}$.

Equations (\ref{pz}) and (\ref{n})  
should be integrated along the group
velocity characteristics $ \partial z / \partial t = v_g$. 
Assuming that all quantities depend on the variable $\xi= z- v_g t$,  
equations (\ref{pz}) and (\ref{n}) become
\ba
&&
- v_g \partial_{\xi} p_z + \partial_{\xi} \gamma=0
\label{pz1}
\\
&&
 \partial_{\xi} \ln n = { \partial_{\xi} v_z \over v_g - v_z}
\label{n1}
\ea
For nonlinear waves, 
the group velocity $v_g$ depends on the Lorentz factor of the
particle flow and on the local plasma density,
  $v_g(n, \gamma )$.
The general solution of the system (\ref{pz1}-\ref{n1})
 for varying $v_g(n, \gamma )$ seems to be untractable,
so we first analytically  illustrate 
the overturn for the case  of constant group velocity
and later resort to numerical integration for the fully nonlinear case. 

Integrating  (\ref{pz1})  along the trajectories $v_g = {\rm const}$,
we find (Clemmow 1974)
\ba
&&
v_z = \frac{{n_g}\,\left( 1 + {{{\nu }_0}}^2 + {\sqrt{1 - \left( -1 + {{n_g}}^2 \right) \,{{{\nu }_0}}^2}} \right) }
  {1 + {{n_g}}^2 + {{{\nu }_0}}^2}
\nn
&&
\gamma =
 \frac{{{n_g}}^2 + {\sqrt{1 - \left( -1 + {{n_g}}^2 \right) 
\,{{{\nu }_0}}^2}}}{ {{n_g}}^2 -1 }
\label{G}
\ea
where $n_g = 1/ v_g$ and $ \nu_0$ is the  dimensionless amplitude of the
wave.
Equation (\ref{G})  shows that  for the intensity of the wave above
the threshold value
$\nu_0 = v_g / \sqrt{1-v_g^2}$ 
(at which point $v_g = v_z = \frac{{{\nu }_0}}{{\sqrt{1 + {{{\nu }_0}}^2}}}$)
 the Lorentz factor
 $\gamma$ becomes undetermined. This implies that hydrodynamic
description of the medium is no longer valid: the waves overturn 
(Dawson 1959, Lembege \& Dawson 1987; Lembege \& Dawson 1989).
During wave overturn, 
a mixing of various parts of the wave occurs, which 
destroys its oscillatory structure.
Note that in the overturning regions the singularity in the density
(\ref{n1}) at this approximation is marks the regime of strong dissipation,
where the bulk motion is randomized.  It is thus  
a regime of particle acceleration.

\begin{figure}[h]
\psfig{file=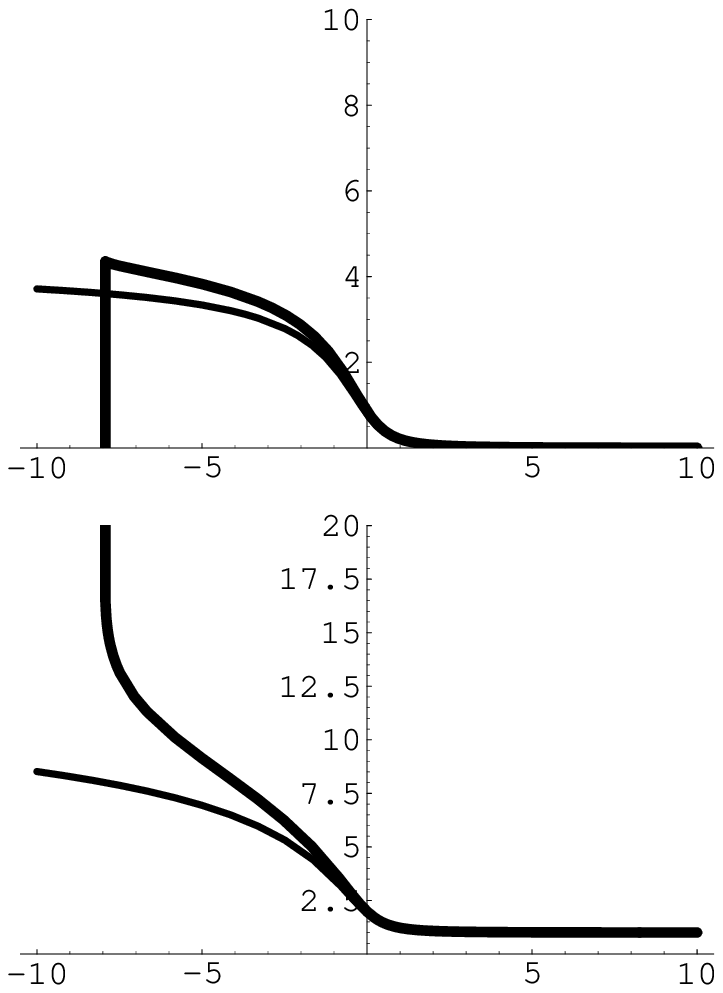,width=10cm}
\caption{ Profiles of (a) parallel momenta and (b) density in a wave. The profile of the
wave is  $ a= \nu_0 \arctan( x- v_g t)$ with $  v_g = 1/\sqrt{ 1+ \mu n/\gamma}$. Density
parameter $\mu =0.1$. Thin lines correspond to subcritical amplitude $\nu_0=2.4$,
thick line correspond to overcritial  amplitude $\nu_0=2.5$. }
\label{F}
\end{figure}

To further illustrate the wave overturn, we numerically integrate 
(\ref{pz}) and (\ref{n})  
for  a given  profile of the incoming wave (Fig (\ref{F})). 

The overturn of strong nonlinear waves is a well 
known phenomena in fiber optics 
where optical shocks (sometimes called kink solutions) form 
when a narrow intense laser pulse propagating
though a nonlinear medium 
(e.g. Agrawal 1989; Rothenberg \& Grishkowsky 1989: 
Trippenbach \& Band  1998). 
Unlike relativistically strong LAEMW, the typical nonlinear intensities of the 
short laser pulses are small and usually a nonlinear 
Schr\"odinger equation
is used to consider wave breaking (Agrawal \& Headly 1992).

In an electron-ion plasma, the ponderomotive force felt by ions and 
electrons is different, resulting in different accelerations and the 
creation of large unstable currents.
Generally, electron-positron plasmas are 
more stable to parametric and modulational
instabilities than ion electron plasmas 
since some of the nonlinear 3rd order currents cancel out.
 However for a  magnetized pair plasma 
the ponderomotive force  acting on electrons and positrons 
is different. Thus if the outflowing plasma of GRBs
has a large ion component or large nonoscillating magnetic fields, then 
the propagation of LAEMWs will be significantly more unstable.

\section{Emission in a random EM field}

\subsection{General remarks}

In this section we consider  generation of 
X-ray and gamma-ray emission  from the 
acceleration and 
propagation of relativistic electrons/positrons 
through relativistically strong electromagnetic 
turbulence. Acceleration of particles is due to 
random alignments of the electric field along the particle's velocity. 
Radiation, on the other hand, is due to the 
 random components of the  magnetic and  
electric field  perpendicular to the particle's velocity. 

The typical electric field  $ E \sim B $ produced near the transition 
radius is orders of magnitude  larger than the Dreicer field
in the mildly relativistic plasma with critical density $n_{GJ}$:
$E_D \sim {e \over r_s^2} \sim \left( { r_e \Omega \over c} \right) B$.
For  electric  fields larger than Dreicer field, Coulomb collisions between 
electrons can be neglected and they are freely accelerated by the field.

The  spectrum of the plasma  particles will be determined by the competition
between acceleration and emission in the stochastic EM fields.
If these two process balance, a stationary spectrum will be reached.
In this case the  energy of particles would be 
radiation limited and only  weakly dependent   on the 
details of the acceleration mechanism.  

Acceleration of particles may proceed in two regimes which we refer to
as ``maximally efficient'' and ``stochastic acceleration'' respectively. 
When the typical scale of the field fluctuations is larger than the 
typical radiative loss length, then the particle
acceleration can be thought of as acceleration in an almost constant
DC electric field and thus will be maximally efficient.
If, on the other hand,  the typical radiative length
 is much larger than the coherence length of the field fluctuation, then
the particle will encounter multiple electric field fluctuations
and reversals before attaining steady state energy.
Here the acceleration is stochastic. 
We expect that electromagnetic turbulence to be broadband, extending
over scales from the
skin depth up to the size of the turbulent region.
The relative importance of the maximally efficient and stochastic acceleration
 types depends on the (unknown) spectrum of the strongly relativistic 
turbulence.

Emission from a relativistic particle in
random $E$ and $B$ fields may be qualitatively related 
to emission from an electron in a random $B$ field alone.
To see this we note that the acceleration due to the 
electric field along the particle's 
 velocity ${\bf E}  \parallel {\bf v} $ 
is much less (by a factor $\sim \gamma^2$)
than the  acceleration due to transverse component of the electric 
field ${\bf E}_{\perp} $.
(Landau \& Lifshitz 1975). The component of the electric field ${\bf E}_{\perp} $
  transverse  to the particle's velocity
 acts similarly to a ${\bf B} $ field.
 Thus the emission of an electron
in random $E$ field will be very similar to the emission of an electron 
in effective magnetic field of the same magnitude.
Moreover, in a relativistic plasma  the random electric and
magnetic fields  should be of the same order.

The particle 
emission properties in turbulent magnetic and electric fields  also depend on
the typical scales of field fluctuations. 
Relativistically moving particles emit most of the radiation along the
particle's velocity within 
an angle $\delta \theta \sim 1/\gamma$. The resulting
spectrum of the emission depends on whether the total deviation angle of the 
particle during its transition through EM field is larger or smaller that 
$\sim 1/\gamma$.
For transverse acceleration, the relevant scale 
is the relativistic Larmor radius
$r_L = c \gamma / \om_{eff}$.
 Perturbations of the  accelerating fields with 
scales larger than $r_L$ produce    deflection 
angles larger than  $\sim 1/\gamma$. 
The resulting emission will be similar to synchrotron
emission. The total spectrum from an 
emitting plasma is then the  average over the emitting volume
of synchrotron spectra.  

On the other hand, for field fluctuations
with wavelengths smaller than the  Larmor radius
 the  deflection during emission is smaller than  $\sim 1/\gamma$.             
The particle moves almost straight along the line and 
experiences high-frequency jitter in the perpendicular direction 
from the random Lorentz force.  Thus  produces ``jitter'' radiation
 (Medvedev 2000, Landau \& Lifshitz II) with a typical frequency
 $\sim \gamma^2 c/\lambda$, where $\lambda$ is the scale length of 
perturbations. The  spectrum is then determined by random 
accelerations of the particle. 
As we show in Section 4.3 however, 
the expected turbulent scales are much larger than the
Larmor radii, so that the emission will resemble synchrotron, not jitter, emission.

\subsection{Order of magnitude estimates}

To illustrate the some of the main points discussed above, 
we first consider  the case when the smallest scale of the field
fluctuations is assumed to be larger than
the radiative loss length, which in turn is assumed to be larger
than the non-relativistic Larmor radius.
In this case, the acceleration and emission result from  
quasistatic  electric and magnetic fields.

Also in this case, the
typical frequency and energy loss of a particle  follows from the analogy
to synchrotron radiation in the an effective
magnetic field $B_{eff} = \sqrt{ <B^2> + <E^2>}$, where
$<B^2>$ and $<E^2> $ are the rms of the fields:  
\be
\om_t \sim \gamma^2 \om_{eff},
\ee
and where $ \om_{eff} = { e \sqrt{ <B^2> + <E^2>} \over m c}$ is the "effective
cyclotron frequency". The total emissivity from a particle is
\be 
{dU \over dt} \sim  \gamma^2 { e^2 \over c} \om_{eff}^2.
\label{dUdt}
\ee
In a steady state, the radiative losses would be exactly compensated by 
acceleration in the fluctuating electromagnetic fields. 
Assuming maximally effective acceleration, we find the energy gain to be 
\be
 dU/dt \sim  {1\over \sqrt{2} }m c^2 \omega_{eff},
\label{U}
\ee
where the factor $1/\sqrt{2} $ is due to the fact that only the 
electric field contributes to acceleration.
Equating Eqns. (\ref{dUdt}) and (\ref{U}) we find
the typical Lorentz factor of the particles to be
\be
\gamma \sim \sqrt{ {r_L \over 2  r_e}} \sim  7 \times 10^4, 
\label{21}
\ee
where $ r_L = c/\om_{eff} $ is the effective Larmor radius and
$r_e = {e^2 /mc^2} $ is the classical electron radius.
The typical  emission energy is then
given by  a combination of fundamental constants only!
\be
\epsilon  \sim  {1 \over \sqrt{2} } { \hbar \, c\over   r_e}\sim 40{\rm MeV}.
\ee
Thus the typical energy of emission of a relativistic 
particle in turbulent EM field 
turns out to be 
independent of the spectral details of the turbulence in the maximally
efficient regime.  The energy of the particles  is radiation limited.

To see this more clearly, 
we compare the energy density in the particles  in a stationary state 
with the equipartition energy density that would be achieved if 
approximately half of the energy of the initial magnetic field
were transformed into particle energy.
The ratio of the magnetic to particle energy density 
may be expressed as a ratio of the
relativistic  Larmor and skin depths
\be
\eta \equiv {U_m \over U_p} = \left( { r_s^{(r)} \over r_L^{(r)} } \right)^2 = {1\over < \gamma >} 
\left( { r_s \over  r_L }\right)^2 >> 1,   
\ee
where 
$ r_s^{(r)} = \sqrt{ <\gamma > } r_s$ and $ r_L ^{(r)}  = < \gamma>  r_L$ are
the relativistic skin depth and the relativistic Larmor radius and the latter
relation follows for the relevant parameter choices.
Since both quantities are properties of the plasma as a whole, we used 
the averaged Lorentz factor of the particles $< \gamma>$ in their definitions.
Had we assumed equipartition, $\eta \sim 1$, 
the average energy   $<\gamma>$ would have been 
\be
<\gamma> \sim \left( { r_s \over  r_L }\right)^2 = 10^9
\label{gg}
\ee
which is much larger than (\ref{21}).
Thus the typical energy density in 
particles is radiation limited and 
is less than the energy density  in EM fields by a factor
$\eta \sim 10^4$. 

The typical synchrotron decay time for a single particle is very short
\be
t_r \sim { c \over r_e \gamma \om_{eff}^2} \approx  5 \times 10^{-8} {\rm sec}
\ee
Yet in a steady state the energy lost by a particle due to radiation is resupplied
by the energy stored in turbulent EM fields. The typical radiation decay times for the
turbulence then will be $\eta$ times longer
\be
\tau_r \sim \eta t_r =5 \times  10^{-4} {\rm sec}
\label{tau_r}
\ee
This radiation decay time 
maybe related to the duration of micro spikes/bursts -  
each micro spike being due to one overturning region.

\subsection{Short turbulence scales: qualitative account of stochastic
acceleration.}

The estimates given above are the upper limits on the emission frequency
and typical energy of particles. They are reached  only if the 
correlation length of the field  fluctuations $l_c$ is of the order of the 
radiative length 
\be
l_r \sim  c { U \over dU/dt} = { r_L^2 \over \gamma r_e } 
\ee
which in the case of maximally efficient acceleration becomes
 $l_r  
r_L \sqrt{{ r_L \over r_e}} $.
If, on the other hand, the correlation length of the
fluctuating electric field is much less than the radiative length, 
$l_c \ll l_r$, the effective accelerating field  will be lower by a factor 
$\sqrt{ l_c / l_r}$, which represents an effective RMS deficit
in the coherence of the particle velocity and accelerating electric field.

Suppose that the acceleration is lower than the  maximum acceleration 
(Eq. (\ref{U}) by a
factor $f< 1$
\be 
{d U \over dt } \sim  f \om_{eff} m c^2.
\ee
Then we would have a typical frequency  of emission 
\be
 \om_t \sim f c/r_e,
\label{nwt} 
\ee 
 a  
typical energy  
\be
<\gamma> = \sqrt{f} { r_L \over r_e},
\label{nga} 
\ee 
and radiative length
\be
l_r \sim {r_L \over \sqrt{f}} \left({r_L \over r_e} \right)^{1/2}
\label{nwlr}. 
\ee
Suppose that the turbulent energy is concentrated near the smallest turbulent
 scale, which  may be approximated 
 by the
relativistic skin depth $ r_s^{(r)} = \sqrt{ <\gamma > } c /\om_p$. 
(This will provide a lower limit to the efficiency
of acceleration.)
The ratio of the radiative length to the smallest scale of turbulence 
$r_s^{(r)}$ then becomes
\be
{ l_r \over  r_s^{(r)} } \sim { 1 \over \gamma^{3/2} } { r_L^2  \over r_s r_e}.
\label{32}
\ee
The efficiency of the acceleration $f$ can be estimated
as $ f \sim \sqrt{ {  r_s^{(r)} \over  l_r }}$. 
Then (\ref{nga}) and (\ref{32}) give
\be
{ l_r \over  r_s^{(r)} } = \left( { r_e r_L \over r_s^2}  \right)^{2/7} = 4\times 10^{-6}.
\ee
Since at the transition radius 
${r_L  \over r_s^2} = { 2 \Omega \over c}$, 
the efficiency of the acceleration $f$ becomes
\be
f = \left( { r_e r_L \over r_s^2}  \right)^{1/7} =  \left( {  2 r_e \Omega \over c} \right)^{1/7} = 
2\times 10^{-3}.
\label{f}
\ee
 The typical energy of emitted photons in this case
\be
\epsilon \sim f  \hbar  {c\over r_e}= 80 \,{\rm  keV}
\label{D}
\ee
and the average energy $<\gamma> = \sqrt{f} r_L/r_e = 5 \times 10^3$.

Relation (\ref{D}) gives the peak energy in the plasma frame. In the laboratory frame
this should be multiplied by the Lorentz factor of the flow.
Since the observed peak energy are around 500 keV, we infer that the bulk
Lorentz factors are relatively small $\Gamma_{bulk} \sim 10$ 
(see also Stern 1999). \footnote{Recall that
in our model the bulk Lorentz factors do not require
 $\Gamma_{bulk} \geq 100$.} The low bulk Lorentz factors may reflect the fact that
MHD acceleration usually is not very effective (Michel 1969).

We would like to stress that Eq. (\ref{f}) was derived by assuming that 
the turbulent energy is
concentrated on the skin depth scale. The presence of larger scale 
electric fields would increase the efficiency factor $f$, so
(\ref{f}) is a lower limit.
The weak dependence 
\footnote{There are three free parameters:$\Omega, \, B_0$ and multiplicity $\lambda$} 
of this lower limit on the 
magnetized rotator's spin $\Omega$,  
its independence of the rotator's magnetic field and thus independence
of the local  plasma density parameter, $\lambda$, are
interesting properties of our model. 
Said another way, when this lower limit on $f$ applies,
the energy of the emitted photons is 
independent of two out of three free parameters of the model
($B_0$ and $\lambda$), and only 
very weakly dependent on the third $ \Omega^{1/7}$.

Next we confirm our earlier assumption that the emission resembles 
synchrotron emission not  "jitter" emission. 
The typical  nonrelativistic 
 Larmor radius $r_L \sim 10^{-3} {\rm cm}$ 
is much less than the smallest turbulent scale (given by the    
the relativistic skin depth $r_s ^{(r)} = c \sqrt{<\gamma>} / \om_p  \sim 10^3 {\rm cm}$,
$<\gamma> = 5  \times 10^3 $.
Thus the   particles  will produce synchrotron-type
radiation in a spatially varying magnetic field.

We can also verify that that the plasma in optically thin to Thompson
scattering ($\tau_T = \sigma_T n r_t \sim 10^{-3}$) and to synchrotron selfabsorbtion
at peak frequency (Pacholczyk Eq. 3.45).

\subsection{Spectrum of turbulence}

The actual spectrum of the relativisticly 
strong EM plasma turbulence has not  yet be determined,
so we resort to general arguments.  
One possibility is a power-law  Kolmogorov-type spectrum in 
with dominant turbulent energy on large scales, transferred 
to small scales. The problem is that
relativistically strong EM turbulence may not satisfy the  following necessary 
conditions required for by 
 Kolmogorov-type turbulence: (i)  the turbulent cascades should be
 local in $k$ space and  (ii) there  should be  an inertial range in $k$ 
space where resistive damping is weak. 
Both these conditions may not be satisfied:
strong turbulence may be nonlocal and damping of turbulent energy 
by plasma particles maybe important.
\footnote{
Dettmann \& Frankel (1996) argued that in relativistic 
hydrodynamic turbulence $\Gamma \propto r$, where $r$ is the scale parameter.
if this is true, then ${\cal{E}}_k  \propto k^{-4}$ (compare with Kolmogorov 
${\cal{E}}_k \propto k^{-11/3}$. If this scaling  is true the energy is
 equally distributed on all scales  scales.}

Another possibility
 is a Phillips-type spectrum (Phillips  1958).
 Following Phyllips (1958) we can argue that at
 the point of overturn  the profile of the wave becomes a
discontinuous function (discontinuity of  the zeroth order). 
The large wave number limit
of the Fourier transform of such function will be $\propto k^{-1}$. 
Thus we can argue
that the resulting  
spectrum of the EM turbulence will be ${\cal{E}_k} \propto k^{-2}$.
\footnote{
This is different from the breaking of surface waves on deep water, which form
unstable  cusps with a discontinuous derivative 
(discontinuity of the first  order). 
In the case of surface waves the 
spectrum of the turbulence is  $\propto k^{-4}$
(Phillips spectrum).
}

Assuming that ${\cal{E}_k} = C k^{-2}$
 and that the smallest scale of the turbulence is given by the
relativistic skin depth, we find
\be
 {\cal{E}} = \int d^3 k {\cal{E}_k}  \sim C \int dk  =   C  k_{max} = C {1\over r_s}.
\ee
In this case, the maximum power 
per range of $d k  $ is $\propto k$. Most of the
turbulent energy is indeed concentrated at the  smallest 
turbulent scales $ \sim r_s$.
For Kolmogorov-type turbulence, $ {\cal{E}_k} \propto k^{-11/3}$ so that most of the 
turbulent energy is concentrated at the large scales.

The turbulent plasma may also produce emission due to nonlinear conversion 
of the turbulent EM fields into escaping radiation. 
The typical frequencies will be of the order of the plasma frequency, 
$\om_p \sim 10^9 {\rm sec^{-1}}$, falling into the radio wave band.
 Another possibility is the production of radio waves by the overturning wave itself -
an analogue of the supercontinuum generation in optical shocks (e.g. Zozulya et al. 1999).

\subsection{Temporal  evolution and spectra}

The temporal evolution of the emission from the foaming region depends on the 
onset and 
 temporal behavior of the electromagnetic turbulence.
In the initial stages of the wave breaking the energy of the initially 
coherent
EM wave may be  deposited at large scales, comparable to the size of the breaking region
($\sim 10^{7} cm$). 
After the initial energy injection, 
the energy cascades to smaller
scales   through turbulent interaction,
 reaching the smallest scale of the turbulence, the skin depth.

This turbulent cascade may or may not reach a quasistationary state. The typical time
for the development of the turbulent cascade maybe as large as 
 the overturn time of the largest turbulent 
eddies, $\sim 1 $ sec,  if the turbulent cascade is local in $k$ space, 
or as small as the typical microscopic interaction time,  $\sim 1/\om_p = 10^{-8}$ sec,
 if the turbulent cascade is nonlocal in $k$ space. 
 The properties of the relativistically strong plasma turbulence have not been investigated yet, so
 we leave the question of the  quasistationary state open
and discuss observational features of both cases.

If it takes longer than the plasma synchrotron cooling time
for the turbulent energy to cascade to the smallest scales, 
then the quasistationary turbulent spectrum may not be reached. 
In this case we expect that the observed 
emission from such non-stationary turbulence will 
be almost time independent until 
(and if) the turbulence cascades to the radiation length scale. 
The reason is that initially the coherence size of the 
electric fields is much larger than the radiation length so 
the initial particle acceleration
is maximally efficient. Particles are accelerated to roughly the same
energy (since the energy is radiation, not acceleration, limited), 
so that the emission spectrum is quasistationary and 
similar to the synchrotron spectrum 
from a mono-energetic particle distribution. 
When the  turbulent cascade reaches the 
radiation length, the acceleration ceases to be maximally efficient. 
The particles are then accelerated in stochastic electric field, so that their
spectrum will be  power law with time dependent diffusion coefficient.
The diffusion coefficient will be decreasing with time, acceleration will be 
less and less efficient and correspondingly the
observed spectrum will be softening with time. 

In the limit of a very slow  energy cascade, the radiation losses
may drain the energy contained in the EM turbulence before the typical 
turbulent scale reaches the radiation length.
 In this case the spectrum of the particles and of the
emitted radiation will resemble the spectra of synchrotron 
cooling sources.

If on the other hand the  turbulent energy  cascades to small scales  faster 
than the plasma synchrotron cooling time, 
a quasi stationary distribution of electrons will be reached. 
Acceleration due to small scale electric fields will be stochastic
while acceleration  due to  large scale electric fields  will be of  DC type.
The particle acceleration will depended on the spectrum of the EM turbulence. 
If most of the energy is concentrated 
at large scales, as in Kolmogorov turbulence, then acceleration
will be maximally efficient and the spectrum of particles will be approximately
monoenergetic. If, on the other hand, most of the energy is concentrated 
 at small scales, the acceleration will be stochastic.

The observed spectra of GRBs will depend on the distribution function
of emitting electrons, which, in turn, depends on several unknown 
parameters like the prevailing type of acceleration
(stochastic or DC-type), proper boundary and initial conditions 
(continuos ejection of particles at low or  high energies
or "no flux" equilibrium states (Melrose 69, 71, 1980,  Tademaru 71, Park and Petrosyan, 
 Katz 1994)),
relative importance of escape of highly energetic particles in the long duration bursts.
The stationary  "no flux" equilibrium (Tademaru 71) will produce
a relativistic Maxwellian distribution 
$f(\gamma) \propto \gamma^{\alpha} e^{-\gamma/\gamma_0} $, with 
$\alpha >2$, while "reflecting boundaries" condition of  Melrose (69, 71, 1980)
will produce power law at $\gamma > \gamma_0$.
DC-type acceleration will tend to produce  strongly peaked (monoenergetic) distribution,
while stochastic acceleration will lead to spectral broadening.
In addition, 
if particle escape is important, the resulting spectra will be powerlaw-like, depending on the
energy dependence of the escape probability (Melrose 1980).
\footnote{
In this respect we also note that the "plasma turbulent reactor" (refs**), which bears 
 some similarity to our model problem, produces a power-law spectrum $p^{-3}$,
consistent with the observed spectral index $-2$.}
This variety of possible spectra may serve well to explain the unusual variety of 
GRB's spectra, which are often powerlaws at small and large energies
and sharply
peaked near the spectral break (Band et al. 1993).
We leave a more
detailed consideration of spectral properties for a subsequent work.

\section{Regeneration of magnetic fields and afterglows}

The interface between the accelerated plasma 
and the ambient medium should lead to the formation of an
external shock at large distances from the magnetized rotator.
This external shock could correspond to that inferred to
be responsible for GRB afterglows (c.f. Meszaros \& Rees 1993;
Sari et al. 1998; Piran 1999; Wijers \& Galama 1998;\cite{FWK99,GPS99}),  
 at large distances. To fit the afterglows with the 
synchrotron emission of relativistic particles large magnetic fields, of 
the order of equipartition 
are required. Generation of such large magnetic fields have presented 
an unsolved problem (\cite{Gruz99}) in the blast wave model. 

In the PFDO model there is a  possibility of regeneration
of  magnetic fields due to interaction of LAEMW with plasma.
It has been known
that interaction of LAEMW in the form of 
lasers with plasma produces both large and small
scale magnetic field (e.g., Stamper et al. 1975; 1978, Sudan 1993,
 Max 1980). 
 Most commonly,  magnetic fields are generated   on skin depth scales
(Pegoraro et al. 1997). 
The typical regenerated magnetic fields may be as large as initial magnetic field
$B \sim 10^6$ G. It has been argued that such small scale quasistationary 
magnetic fields may explain the polarization observed in the afterglows 
(Medvedev 2000).
Alternatively, 
 the  magnetic fields may
be also  subject to  coalecence
 instability (Finn \& Kaw 1977, Sakai et al. 199),
 in which randomly
oriented magnetic islands merge forming  large scale structures.

There is also a possibility of generating large scale magnetic fields, with
correlation length of the order of the size of inhomogeneities of the
"of the laser beam" by inverse Faraday effect (Berezhiani et al. 1997),
which may generate fields of the order 
\be
B \sim { \gamma_0  m c^2 \over e R_{\perp}}  
\ee
where $ \gamma_0 \sim \nu_0 $ is the typical Lorentz factor of particles accelerated by a
EM pulse with a transverse dimension $R _{\perp}$, 
in our case the transverse size of the foaming region.

We mention these possibilities here as a flag for future work,
and to highlight the need to further investigate possible analogies between
field amplification in GRB and that of laser driven plasmas.
The connection is that both plasma systems may be driven by LAEMW.

\section{Conclusion}

The two main points of this work may be summarized as follow:
(i) Poynting flux dominated outflows, which so far has been invoked only in the
context of central engines, possesses internal instabilities which may also 
explain radiation generation, (ii) 
emission from stochastically accelerated
particles  in turbulent electromagnetic fields, 
when electric fields are  as important as magnetic 
fields in the acceleration and radiation process, is a viable mechanism
for the GRB emission.  

Our approach may also naturally explain 
the result that the peak of GRB 
emission varies only over a small range of value from burst to burst. 
(Brainerd 1994, 1997).  For our maximally efficient 
mode of acceleration, the peak, when
boosted into observer frame, becomes
 $ \epsilon_{max} \sim  f \hbar c/r_e  \Gamma_{\rm bulk}$. 
If this maximally efficient regime were operating,
the bulk gamma-factor $\Gamma_{\rm bulk}$ would not be  much larger than 10. 
The location of this peak is  a very weak 
function of the parameters of the underlying rotator, 
 $f \sim \Omega ^{1/7} $ and is independent of the progenitor's
magnetic field. 

This can be compared  to phenomenological 
 internal shock models in which the peak energy is proportional to
$\gamma_{min}^2 \om_B \Gamma_{bulk}$, where $\gamma_{min}$ is the low energy 
cutoff to the electron power law spectrum. All three quantities here are
usually taken to be free parameters. 
It seems unlikely for their combination to be 
almost constant from burst to  burst, 
and within each pulse of a given burst 
unless some physics dictates this to be the case.
We have tried to add some of this physics.
In our approach the remaining free  parameter is the Lorentz factor of the 
bulk flow.

PFDOs may also resolve the problem of the magnetic field generation since
 in the  Poynting flux driven  outflows the 
large magnetic fields are supplied by the source. These fields
do not necessarily have to be generated in the external shock for the
afterglow since the shock is a current sheet through which outflow and
ambient particles mix (Smolsky \& Usov 2000).  At the same time, however 
there exists an unexplored analogy between field generation mechanisms in laser driven
plasmas and GRB outflows that will have to be understood to determine
the scale and structure of the outlfow  magnetic field.

Other observational properties of the GRBs that may be explained in our
framework:
(i) 
GRBs show no correlation between the spectra and other micropulse characteristics 
(e.g., Lee \& Bloom 2000) - 
this is a direct consequence of the turbulent EM acceleration which produces 
photons with the frequency $\sim c/r_e$;
(ii)
the fact that pulses peak earlier at high frequencies and that bursts have shorter 
duration  at higher energies  may be due to 
the initial development of the turbulence: if the turbulence develops from large scales to small scales
then initially the acceleration may be  more effective since it
is due to large scale electric fields
(with coherence length larger than the radiative length). In this case  initially particles are
accelerated to the limiting energies $\gamma \sim \sqrt{r_L /r_e}$ while later, when
the  coherence length of the turbulence becomes smaller than the  radiative length,
the particle distributions soften  and radiation spectra emanates 
at lower frequencies. 
The typical time for such a cascade should be of the order of the "vortex overturn time"
on the largest scale of the turbulence, which may be on the order of seconds.
(iii) the absence of correlation between the pulses and overall burst characteristics, 
interpreted as arising from random and independent  emission episodes (Lee \& Bloom 2000),
is natural in our model since 
each wave overturn happens independently.
(iv)
The temporal characteristics of the microbursts, FREDs,  and the hard-to-soft
spectral evolution is a consequence
of synchrotron cooling of the  reservoir of 
energy released during wave overturn.
(v) composite structure of GRBs (a burst being a sum of many 
 independent  emission events) (Stern \& Svennson 1996) naturally follows from the 
model - each overturning region produces an independent microburst.
(vi)  the average power density spectrum of GRBs is well described as being due to 
selfsimilar turbulent-type process near marginal stability 
(Stern 1999). This is reminicent of the cellular automata model of solar
 reconnecting regions (Lu \& Hamilton 1991) and may be related to the
reconnecting "foaming" regions in our model.

We also would like to point out that the 
 particular mechanism of the wave instability, the wave overturn during 
MHD-wave transition, may not necessarily be the only one.
Other plasma instabilities may contribute to the generation 
of EM turbulence.

\acknowledgements
 We would like to thank Michail Medvedev, Vladimir Usov, Norm Murray,
Roger Blandford and  Andrei Gruzinov for useful comments.

\end{document}